\documentclass[dvips]{acta}
\usepackage{supertabular,lscape,epsfig}
\usepackage{amssymb}
\usepackage{amsmath}
\DeclareSymbolFont{ppa}{OT1}{ppl}{m}{it}
\DeclareMathSymbol{\vv}{\mathalpha}{ppa}{'166}

\newfont{\hb}{rphvb at 10pt}%bezszeryfowe pó³grube
\newfont{\hbo}{rphvbo at 10pt}%bezszeryfowe pó³grube kursywa
\newfont{\bitt}{rptmbi at 12pt}%pó³gruba kursywa (tytu³ artyku³u)
\newfont{\bits}{rptmbi at 11pt}%pó³gruba kursywa (tytu³y rozdzia³ów)

\SetPages{1}{1}

\SetVol{64}{2014}

\begin{document}

%Zwarte naglowki, jeden wiersz
\newcommand{\TabCapp}[2]{\begin{center}\parbox[t]{#1}{\centerline{
  \small {\spaceskip 2pt plus 1pt minus 1pt T a b l e}
  \refstepcounter{table}\thetable}
  \vskip2mm
  \centerline{\footnotesize #2}}
  \vskip3mm
\end{center}}

%Zwarte naglowki, dwa wiersze
\newcommand{\TTabCap}[3]{\begin{center}\parbox[t]{#1}{\centerline{
  \small {\spaceskip 2pt plus 1pt minus 1pt T a b l e}
  \refstepcounter{table}\thetable}
  \vskip2mm
  \centerline{\footnotesize #2}
  \centerline{\footnotesize #3}}
  \vskip1mm
\end{center}}

%Zwarte naglowki, jeden wiersz
\newcommand{\MakeTableSepp}[4]{\begin{table}[p]\TabCapp{#2}{#3}
  \begin{center} \TableFont \begin{tabular}{#1} #4
  \end{tabular}\end{center}\end{table}}

%Zwarte naglowki, jeden wiersz
\newcommand{\MakeTableee}[4]{\begin{table}[htb]\TabCapp{#2}{#3}
  \begin{center} \TableFont \begin{tabular}{#1} #4
  \end{tabular}\end{center}\end{table}}

%Zwarte naglowki, dwa wiersze
\newcommand{\MakeTablee}[5]{\begin{table}[htb]\TTabCap{#2}{#3}{#4}
  \begin{center} \TableFont \begin{tabular}{#1} #5
  \end{tabular}\end{center}\end{table}}

\newfont{\bb}{ptmbi8t at 12pt}
\newfont{\bbb}{cmbxti10}
\newfont{\bbbb}{cmbxti10 at 9pt}
\newcommand{\uprule}{\rule{0pt}{2.5ex}}
\newcommand{\douprule}{\rule[-2ex]{0pt}{4.5ex}}
\newcommand{\dorule}{\rule[-2ex]{0pt}{2ex}}
\def\thefootnote{\fnsymbol{footnote}}
\begin{Titlepage}
\Title{Over 38\,000 RR~Lyrae Stars in the OGLE Galactic Bulge
Fields\footnote{Based on observations obtained with the 1.3-m Warsaw
telescope at the Las Campanas Observatory of the Carnegie Institution
for Science.}}
\Author{I.~~S~o~s~z~y~ñ~s~k~i$^1$,~~
A.~~U~d~a~l~s~k~i$^1$,~~
M.\,K.~~S~z~y~m~a~ñ~s~k~i$^1$,~~
P.~~P~i~e~t~r~u~k~o~w~i~c~z$^1$,\\
P.~~M~r~ó~z$^1$,~~
J.~~S~k~o~w~r~o~n$^1$,~~
S.~~K~o~z~³~o~w~s~k~i$^1$,~~
R.~~P~o~l~e~s~k~i$^{1,2}$,~~
D.~~S~k~o~w~r~o~n$^1$,\\
G.~~P~i~e~t~r~z~y~ñ~s~k~i$^{1,3}$,~~
\L.~~W~y~r~z~y~k~o~w~s~k~i$^{1,4}$,~~
K.~~U~l~a~c~z~y~k$^1$,~~
and~~M.~~K~u~b~i~a~k$^1$}
{$^1$Warsaw University Observatory, Al.~Ujazdowskie~4, 00-478~Warszawa, Poland\\
e-mail: (soszynsk,udalski)@astrouw.edu.pl\\
$^2$ Department of Astronomy, Ohio State University, 140 W. 18th Ave., Columbus, OH 43210, USA\\
$^3$ Universidad de Concepción, Departamento de Astronomia, Casilla 160--C, Concepción, Chile\\
$^4$ Institute of Astronomy, University of Cambridge, Madingley Road, Cambridge CB3 0HA, UK}
\Received{~}
\end{Titlepage}
\Abstract{We present the most comprehensive picture ever obtained of the
central parts of the Milky Way probed with RR~Lyrae variable stars. This
is a collection of 38\,257 RR~Lyr stars detected over 182 square
degrees monitored photometrically by the Optical Gravitational Lensing
Experiment (OGLE) in the most central regions of the Galactic bulge. The
sample consists of 16\,804 variables found and published by the OGLE
collaboration in 2011 and 21\,453 RR~Lyr stars newly detected in the
photometric databases of the fourth phase of the OGLE survey (OGLE-IV).
93\% of the OGLE-IV variables were previously unknown. The total sample
consists of 27\,258 RRab, 10\,825 RRc, and 174 RRd stars. We provide
OGLE-IV {\it I}- and {\it V}-band light curves of the variables along with
their basic parameters.

About 300 RR~Lyr stars in our collection are plausible members of 15
globular clusters. Among others, we found the first pulsating variables
that may belong to the globular cluster Terzan~1 and the first RRd star in
the globular cluster M54. Our survey also covers the center and
outskirts of the Sagittarius Dwarf Spheroidal Galaxy enabling studies of
the spatial distribution of the old stellar population from this galaxy.

A group of double-mode RR~Lyr stars with period ratios around 0.740 form
a stream in the sky that may be a relic of a cluster or a dwarf galaxy
tidally disrupted by the Milky Way. Three of our RR~Lyr stars
experienced a pulsation mode switching from double-mode to single
fundamental mode or {\it vice versa}. We also present the first known RRd
stars with large-amplitude Blazhko effect.} {Stars: variables: RR~Lyrae
-- Stars: oscillations (including pulsations) -- Stars: Population II --
Galaxy: center -- Galaxies: individual: Sagittarius Dwarf Spheroidal
Galaxy}

\Section{Introduction}

RR~Lyrae stars are an invaluable source of knowledge about the oldest
stellar population. They are numerous, present in various stellar
environments, and are relatively easy to detect due to their short periods,
characteristic light curves and large amplitudes. RR~Lyr stars are standard
candles which makes them important distance indicators and tracers of the
structure of the Milky Way and other galaxies. RR~Lyr variables are
radially pulsating horizontal branch stars, with periods in the range of
0.2--1.0~d. They are classified into three main types according to the mode
in which they pulsate: fundamental-mode RRab stars, first-overtone RRc
stars and double-mode RRd stars.

Based on the observations carried out during the second and the third
phases of the Optical Gravitational Lensing Experiment (OGLE-II and
OGLE-III projects), we have compiled the largest ever collection of
RR~Lyr stars, consisting in total of over 44\,000 objects. It includes
RR~Lyr stars from the Magellanic Clouds (Soszyñski \etal 2009, 2010),
the Galactic bulge (Soszyñski \etal 2011, hereafter S11) and the
Galactic disk (Pietrukowicz \etal 2013).

The OGLE-discovered RR~Lyr stars have been the base of many research
pro\-jects: studies of the structure of the Magellanic Clouds and
Galactic bulge (\eg Sub\-ra\-ma\-niam and Subramanian 2009, Pejcha and
Stanek 2009, Feast \etal 2010, Kapakos and Hatzidimitriou 2012,
Pietrukowicz \etal 2012, D{\'e}k{\'a}ny \etal 2013, Wagner-Kaiser and
Sarajedini 2013, Deb and Singh 2014, Sans Fuentes and De Ridder 2014),
measurements of distances (\eg Haschke \etal 2012, Pietrukowicz \etal
2012), preparation of the reddening maps (\eg Haschke \etal 2011,
Nataf \etal 2013, Wagner-Kaiser and Sarajedini
2013), studies of the Blazhko modulation in the well-sampled OGLE light
curves of RR~Lyr stars (\eg Chen \etal 2013, Welch 2014), applications
as a training set for automated methods of classification of variable
stars (\eg Long \etal 2012). One of the objects included in the OGLE-III
Catalog of Variable Stars -- OGLE-BLG-RRLYR-02792 -- initially
classified as an RR~Lyr star in an eclipsing binary system, turned out
to be a representative of a new class of variable stars, called binary
evolution pulsators (Pietrzyñski \etal 2012, Smolec \etal 2013). Such
stars have masses much smaller than expected for RR~Lyr variables, but
their light curves mimic those of classical pulsators.

OGLE-III RR~Lyr stars were also extensively followed-up in different
wavelength ranges. For example, they served as input list objects for
infrared surveys like the Vista Variables in the V{\i}a L{\'a}ctea (VVV,
D{\'e}k{\'a}ny \etal 2013) or the Carnegie RR~Lyrae Program. Many of
these stars will also be input targets for the large spectroscopic
survey -- APOGEE-II.

In this work -- first of a series presenting variable stars detected
from the observations collected by the fourth phase of the OGLE survey
(OGLE-IV) -- we expand the OGLE collection of RR~Lyr stars in the
central regions of the Galaxy by a factor of more than two and present
the most comprehensive picture ever obtained of the central parts of the
Galactic bulge as seen {\it via} RR~Lyr stars. The number of newly
detected pulsators exceeds 21\,000 and now the whole OGLE sample of
RR~Lyr variables toward the Galactic bulge lines-of-sight consists of
38\,257 stars. This is the largest sample of RR~Lyr stars discovered so
far in one stellar environment. We also provide OGLE-IV light curves of
the RR~Lyr stars detected during the previous stages of the OGLE
project.

The paper is organized as follows. In Section 2 we discuss the observations
used in this study and how these data were reduced. Section 3 presents how
RR~Lyr stars were selected and classified. In Section 4 we compare OGLE-III
and OGLE-IV photometry of RR~Lyr stars. Section 5 describes the structure
of the new catalog of RR~Lyr stars. In Section 6 we estimate the
completeness of our sample. In Section 7 we discuss possible applications
of our collection of RR~Lyr stars and show the most interesting individual
objects. Finally, Section 8 summarizes the conclusions drawn from this
work.

\Section{Observational Data}
The OGLE-IV survey conducts long-term, continuous observations of the
fields toward the Galactic bulge using the 1.3~m Warsaw telescope
located at the Las Campanas Observatory (LCO) in Chile. The observatory
is operated by the Carnegie Institution for Science. OGLE-IV began
observations of the Galactic bulge in March 2010 and regular monitoring
has been continued up to now. The latest observations used in this study
were collected in October 2013, with the exception of several fields
with a relatively small number of points, for which we used observations
collected up to September 2014.

During the OGLE-IV survey the Warsaw telescope is equipped with a
32-detector mosaic CCD camera covering a field of about 1.4 square
degrees in one image. In total, 121 OGLE-IV fields toward the Galactic
center were searched for RR~Lyr stars (\cf the OGLE sky coverage maps on
the OGLE project WWW page: {\it http://ogle.astrouw.edu.pl}). Together
with the OGLE-II and OGLE-III fields analyzed by S11, the bulge area of
182 square degrees was covered. Because the OGLE-IV observing strategy
has been optimized to detect and study microlensing events hosting
exoplanets, the number of collected epochs varies significantly from
field to field -- from about 80 to over eight thousand. Most of the
observations were made through the Cousins {\it I}-band filter with a
standard integration time of 100~s. Additionally from ten to a few
dozen epochs, depending on the field, were also collected in the
Johnson {\it V}-band with the exposure time of 150~s for color
information. The {\it I}-band magnitudes in the OGLE-IV databases range
from about 13~mag to 20.5~mag. 

The photometric reductions were performed using the OGLE real time
photometric pipeline (Udalski 2003) implemented for OGLE-IV observing
set-up. The pipeline is based on Difference Image Analysis technique
(DIA, Alard and Lupton 1998, Wo¼niak 2000). The instrumental photometry was
then stored in the standard OGLE databases.

The final photometry of selected variables was calibrated to the
standard {\it VI} system. Generally, the OGLE-IV photometry has been
tied to precisely calibrated OGLE-III Photometric Maps of the Galactic
Bulge (Szymañski \etal 2011). These calibrations are based on
observations collected on hundreds of photometric nights. The accuracy
of the zero points of OGLE-IV photometry in the Galactic bulge is at
the level of 0.02~mag.

\vspace*{9pt}
\Section{Selection and Classification of RR~Lyr Stars}
In order to detect RR~Lyr variables, we performed an extensive period search
for all discrete sources observed by OGLE-IV toward the Galactic bulge. For
nearly 400 million {\it I}-band light curves with more than 30 data points,
we calculated Fourier amplitude spectra in the frequency range from 0 to
24~d$^{-1}$. For each star we recorded a period corresponding to the
highest peak in the spectrum along with its amplitude and signal-to-noise
ratio.

The identification of RR~Lyr stars was performed in two ways, but in
both approaches the final decision was made based on the visual
inspection of the light curves. In the first method, each light curve
with a period between 0.2 and 1.0~d was fitted with a Fourier cosine
series and the Fourier parameters $R_{21}$, $\phi_{21}$, $R_{31}$,
$\phi_{31}$ (Simon and Lee 1981) were derived. We visually checked the
light curves with the Fourier coefficients and amplitudes typical for
RR~Lyr stars. The second algorithm was based on the template light
curve fitting. First, we prepared templates of typical light curves of
RRab and RRc stars. Then we correlated the real observations with the
templates and visually examined the best fit light curves. 

If the same star was identified twice, in two overlapping fields, only
one entry in our final RR~Lyr star list was left. We have chosen the
detection with larger number of data points in the light curve. Objects
with double detections were used to estimate the completeness of the
catalog (see Section~6).

In this way we found 21\,453 RR~Lyr stars that were not recorded by the
previous stages of the OGLE survey. The sample was divided into three
groups: RRab, RRc and RRd stars. In most cases our classification was
based on the morphology of the light curves. RRd stars were found based
on their characteristic period ratios. We did our best to minimize
contamination of our RR~Lyr sample from other types of variables. RRab
stars are usually easy to distinguish from non-pulsating variable stars
(although sometimes spotted variables may mimic pulsating stars). On the
other hand, RRc stars have much more symmetric light curves and can be
easily confused with close binary systems or rotating stars. In the OGLE
databases we found at least a few thousand sinusoidal light curves with
periods in the range of 0.2 -- 0.5 days. Many of these objects may be
first-overtone RR~Lyr stars, but they are not included in our sample. We
required at least a small asymmetry of the light curves to classify
stars as RRc variables. A number of multi-mode $\delta$~Sct stars
contaminating our RR~Lyr sample were filtered out based on their
position in the Petersen diagram (period ratio \vs period diagram).
However, despite our efforts, we expect that a limited number of
single-mode $\delta$~Sct stars and variable stars of other types may be
hidden among of RR~Lyr stars in our collection. Some questionable cases
are flagged as ``uncertain'' in the Remarks file.

\Section{Comparison with the OGLE-III Catalog}

Despite the fact that OGLE-IV fields cover most of the OGLE-II and
OGLE-III fields in the Galactic bulge, the vast majority of the newly
detected RR~Lyr stars (more than 97\%) were found in the regions outside
the previously searched area. This confirms high completeness of the
OGLE catalogs of variable stars. One should be aware that due to the
technical gaps between CCD detectors of the OGLE-IV mosaic camera which
have been only partially filled in the reference images of a given field
(being a stack of a few individual somewhat shifted images) a few
percent of the area of each OGLE-IV field was excluded from our search.
However, in the fields overlapping with our previous studies (S11) these
OGLE-IV ``dead zones'' were usually filled by OGLE-III observations.

\begin{figure}[t]
\includegraphics[width=12.7cm]{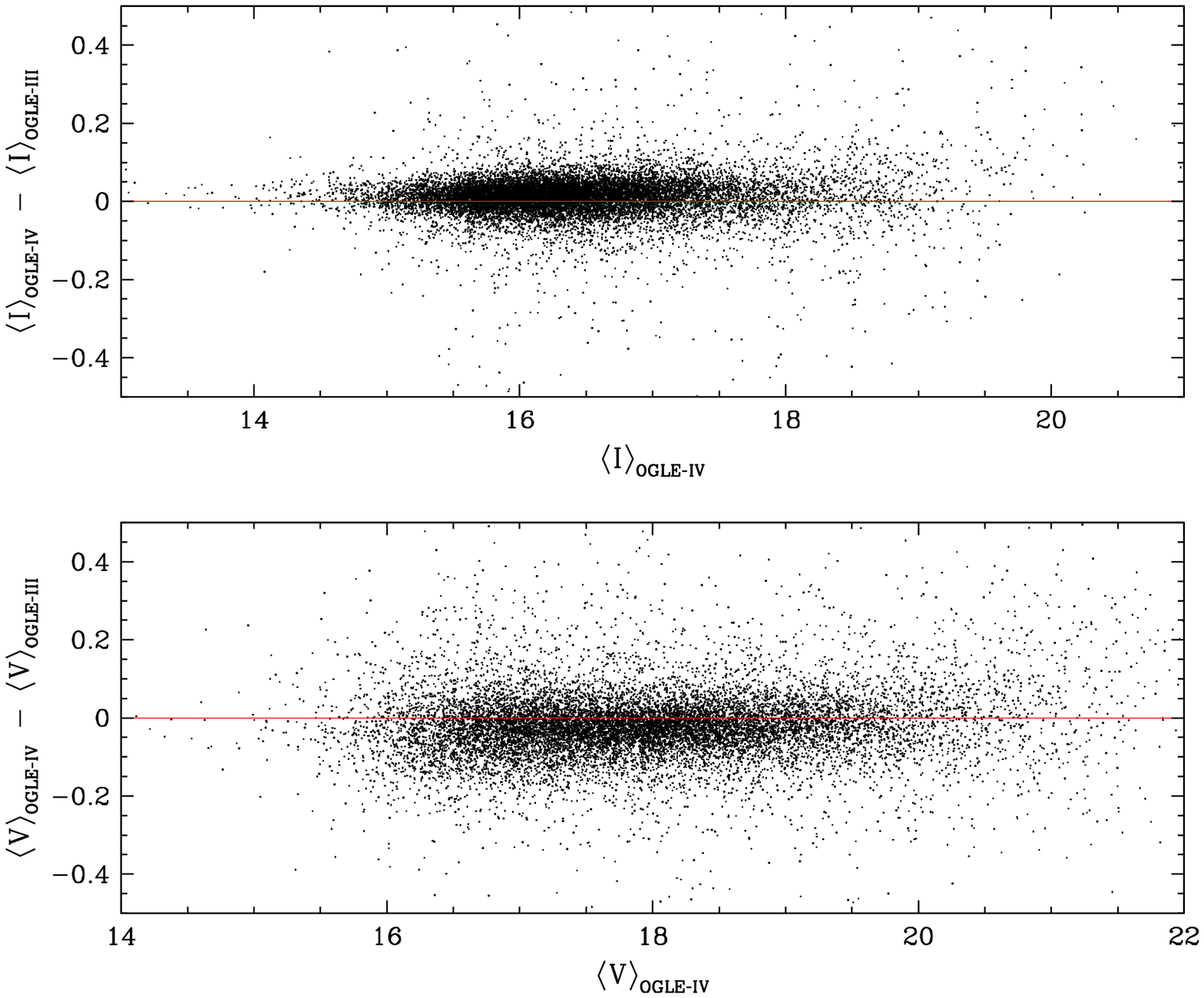}
\FigCap{Comparison between OGLE-III and OGLE-IV {\it I}-band ({\it upper
panel}) and {\it V}-band ({\it lower panel}) mean magnitudes of RR~Lyr
stars in the Galactic bulge.}
\end{figure}

We extracted and visually inspected OGLE-IV light curves of RR~Lyr stars
from the OGLE-III catalog (S11). We recalculated their periods and other
parameters, such as their mean magnitudes, amplitudes and Fourier coefficients.
During the inspection we reclassified 32 objects from S11 as classical
Cepheids, eclipsing binaries, spotted variables or other (usually
unknown) types of variable stars. Table~1 lists all these stars along
with their new classification. We removed these objects from the present
version of our RR~Lyr sample. The binary evolution pulsator
OGLE-BLG-RRLYR-02792 (Pietrzyñski \etal 2012) is not a classical RR~Lyr
star, but we decided to leave it on the list, so that the astronomical
community has an access to the current photometry of this unique object.

\MakeTableee{
l@{\hspace{8pt}} c@{\hspace{6pt}} | l@{\hspace{8pt}} c@{\hspace{6pt}}}{12.5cm}
{Reclassified objects from the OGLE-III Catalog of RR~Lyr stars (S11).}
{\hline \noalign{\vskip3pt}
\multicolumn{1}{c}{ID} & New            & \multicolumn{1}{c}{ID} & New            \\
                       & classification &                        & classification \\
\noalign{\vskip3pt}
\hline
\noalign{\vskip3pt}
OGLE-BLG-RRLYR-01594 & Eclipsing & OGLE-BLG-RRLYR-09947 & Other     \\
OGLE-BLG-RRLYR-01929 & Cepheid   & OGLE-BLG-RRLYR-10243 & Spotted   \\
OGLE-BLG-RRLYR-02159 & Other     & OGLE-BLG-RRLYR-10938 & Cepheid   \\
OGLE-BLG-RRLYR-02251 & Cepheid   & OGLE-BLG-RRLYR-11514 & Other     \\
OGLE-BLG-RRLYR-02674 & Other     & OGLE-BLG-RRLYR-11616 & Other     \\
OGLE-BLG-RRLYR-02676 & Other     & OGLE-BLG-RRLYR-11907 & Other     \\
OGLE-BLG-RRLYR-02974 & Other     & OGLE-BLG-RRLYR-11947 & Other     \\
OGLE-BLG-RRLYR-03385 & Eclipsing & OGLE-BLG-RRLYR-13883 & Other     \\
OGLE-BLG-RRLYR-03439 & Other     & OGLE-BLG-RRLYR-14003 & Eclipsing \\
OGLE-BLG-RRLYR-03929 & Other     & OGLE-BLG-RRLYR-14084 & Eclipsing \\
OGLE-BLG-RRLYR-04475 & Other     & OGLE-BLG-RRLYR-14183 & Other     \\
OGLE-BLG-RRLYR-04864 & Cepheid   & OGLE-BLG-RRLYR-14754 & Spotted   \\
OGLE-BLG-RRLYR-04897 & Other     & OGLE-BLG-RRLYR-15159 & Other     \\
OGLE-BLG-RRLYR-08477 & Other     & OGLE-BLG-RRLYR-15476 & Eclipsing \\
OGLE-BLG-RRLYR-09105 & Cepheid   & OGLE-BLG-RRLYR-16120 & Other     \\
OGLE-BLG-RRLYR-09326 & Other     & OGLE-BLG-RRLYR-16833 & Other     \\
\noalign{\vskip3pt}
\hline}

For several RR~Lyr stars we corrected their mode of pulsation. At least
three variables evidently switched their pulsation modes, from RRd to
RRab or {\it vice versa} (see Section~7.4). In other cases, the OGLE-III
light curves were affected by a small number of points, which resulted
in erroneous classifications by S11. Four other RR~Lyr stars in the
OGLE-III catalog had incorrect periods -- one-day aliases of the real
ones.

The OGLE-III and OGLE-IV light curves have been obtained with different
instrumental configurations, in particular with different filters and
CCD detectors. Although both instrumental systems were transformed to the
standard photometric system, the systematic uncertainties of the
calibration zero point may reach 0.02~mag. In Fig.~1 we compare {\it I}-
and {\it V}-band mean luminosities of RR~Lyr stars derived from the
OGLE-III (S11) and OGLE-IV light curves. The agreement between both data
sets for most of the stars is reasonably good. The mean difference
between {\it I}-band magnitudes measured from the OGLE-IV and OGLE-III
photometry is 0.012~mag with the standard deviation of 0.043~mag. There is
also a number of individual variables for which OGLE-III and OGLE-IV
luminosities differ much more considerably. For 6\% of RR~Lyr stars
observed in both phases of the survey,
$|I_{\text{OGLE-IV}}-I_{\text{OGLE-III}}|$ is larger than 0.1~mag,
and for 2\% this difference is larger than 0.2~mag. We found that in
most such cases the offset between magnitudes was caused by crowding and
blending by unresolved stars that randomly affected the reference image
fluxes of the DIA photometry.

In the {\it V}-band there is a systematic difference between OGLE-IV and
OGLE-III datasets at the level of $-0.020$~mag with the standard deviation of
0.079~mag. The offset is well within the expected uncertainty of the
photometric calibrations. The larger scatter of the points in the lower
panel of Fig.~1 can be explained by a much smaller number of observations
in the {\it V}-band than in {\it I}-band, which affected the accuracy of
the derived mean luminosities. The OGLE-III {\it V}-band light curves
contained typically only a few points. In the OGLE-IV databases the median
number of the {\it V}-band epochs is 22, so we expect that the accuracy of
the calculated mean magnitudes is on average much better.

\Section{Catalog of RR~Lyr Stars Toward the Galactic Bulge}

The sample of newly detected RR~Lyr stars toward the Galactic bulge was
combined with the OGLE-III catalog (S11). The total number of RR~Lyr
variables in our collection is now 38\,257, including 27\,258 RRab,
10\,825 RRc and 174 RRd stars. There are 32 objects reclassified in the
present investigation (Section~4) that were removed from the collection. We
kept the identifiers of RR~Lyr variables from S11. The identifiers from
OGLE-BLG-RRLYR-00001 to OGLE-BLG-RRLYR-16836 are occupied by the RR~Lyr
stars from the OGLE-III catalog. The newly detected variables are named
starting from OGLE-BLG-RRLYR-16837 to OGLE-BLG-RRLYR-38289 and are ordered by
increasing right ascension.

\begin{figure}[p]
\includegraphics[width=12.8cm]{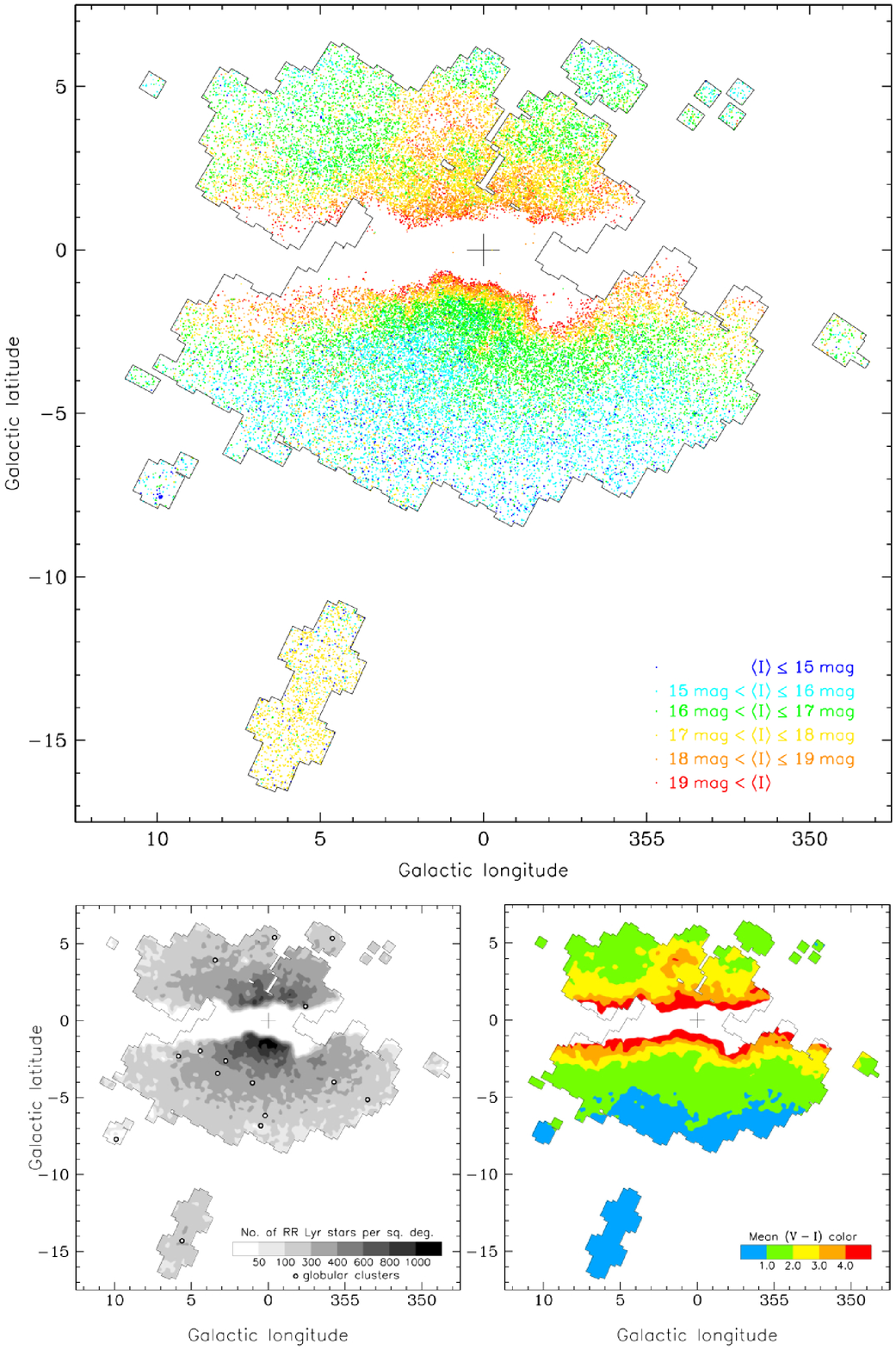}
\FigCap{Spatial distribution of RR~Lyr stars in the OGLE fields toward the
Galactic bulge. In the {\it upper panel} points of different colors
represent individual RR~Lyr stars of different mean magnitudes. {\it Lower
left panel}: surface density map of RR~Lyr stars. White circles indicate
the positions of globular clusters likely hosting RR~Lyr stars. {\it Lower right
panel}: spatial distribution of the mean apparent $(V-I)$ colors of RR~Lyr
stars, that increase toward the Galactic plane due to dust extinction.}
\end{figure}

The whole catalog is available {\it via} anonymous FTP or the WWW site:
\begin{center}
{\it ftp://ftp.astrouw.edu.pl/ogle/ogle4/OCVS/blg/rrlyr/}\\
{\it http://ogle.astrouw.edu.pl/}
\end{center}

The format of the files in the FTP archive follows that of S11. For each
star we provide its identifier, J2000 equatorial coordinates, mode of
pulsation, fields and internal numbers in the OGLE-IV, OGLE-III, and
OGLE-II photometric databases (if available) and the cross-matches with
the International Variable Star Index (VSX, Watson \etal 2006), which is
currently the most complete database of known variable stars.
Observational parameters of the RR~Lyr stars: periods, intensity mean
magnitudes in the {\it I}- and {\it V}-bands, {\it I}-band peak-to-peak
amplitudes, and Fourier coefficients $R_{21}$, $\phi_{21}$, $R_{31}$ and
$\phi_{31}$ derived for the {\it I}-band light curves are listed in
separate files. The periods with uncertainties were found with the {\sc
Tatry} code kindly provided by A.~Schwarzenberg-Czerny (1996).

Note that all the observational parameters were computed using solely the
OGLE-IV photometry collected in the years 2010--2013. This refers also to
the OGLE-III RR~Lyr variables, if only the OGLE-IV light curves were
available. For stars with no OGLE-IV photometry we copied the parameters
from the OGLE-III catalog (S11). We recalculated the periods because many
RR~Lyr stars (especially RRc variables) significantly changed periods in
the time that passed between previous and present phases of the OGLE
project. For strictly periodic variables we recommend to merge OGLE-III and
OGLE-IV light curves and to find the pulsation period using the whole OGLE
photometry spanning up to 17 years. While merging the light curves one
should keep in mind that in individual cases the OGLE-III and OGLE-IV
observations may be somewhat shifted in magnitudes (\cf Fig.~1), so
an additional magnitude alignment may be necessary.

The files with the OGLE-IV time-series photometry of each star are stored
in the {\sf phot/} directory. Finding charts for all stars can be
downloaded from the directory {\sf fcharts/}. These are
$60\arcs\times60\arcs$ subframes of the {\it I}-band DIA reference
images. The file {\sf gc.dat} lists RR~Lyr stars that are candidates for
the globular cluster members. The file {\sf remarks.txt} contains
remarks about some RR~Lyr stars. See the {\sf README} file for more
details.

In Fig.~2 we present the 2-D spatial map of our collection. This is the
most complete view of the central parts of the Galactic
bulge ever obtained with RR~Lyr variables as the stellar structure probes.
Brightness of variables is color-coded. It is clearly seen that the 
RR~Lyr stars become gradually fainter in the lines-of-sight closer to
the Galactic plane. This is obviously due to dramatically increasing
interstellar extinction in the optical bands toward these regions,
moving the brightness of RR~Lyr below the OGLE detection limit. However,
it is worth noting that the empty area around the Galactic plane in
Fig.~2 was also searched for RR~Lyr but no foreground objects were
found. In the future near infrared surveys, \eg the VVV survey
(Minniti \etal 2010, Catelan \etal 2013), may complement the OGLE
picture presented in Fig.~2 in this highly obscured region of the
Galactic plane, unreachable for the OGLE optical survey.

In the isolated region at the Galactic latitudes $b<-10^{\circ}$, most
of the RR~Lyr stars have apparent {\it I}-band luminosities between 17
and 18~mag -- much fainter than expected for the bulge members. This is
because these stars belong to the Sagittarius Dwarf Spheroidal Galaxy
(Sgr dSph) which is located behind the Galactic bulge (see Section 7.2).

Lower panels of Fig.~2 show possible applications of our sample of
RR~Lyr stars. The density map (left panel) traces the structure of the
bulge, while the color map (right panel) reflects the spatial
distribution of the interstellar extinction toward the Milky Way center.
More detailed study of the bulge structure will be presented in
Pietrukowicz \etal (2014, in preparation).

\Section{Completeness of the Sample}

The completeness of our list of RR~Lyr stars in the Galactic bulge
depends on many factors, such as the pulsation modes, brightness of the
stars, amplitudes of variations, positions in the sky and number of
points in the light curves. Undoubtedly, the greatest completeness and
the least contamination is in the regions covered by both, OGLE-III and
OGLE-IV, surveys, since we performed two independent searches for RR~Lyr
stars in this area.

The completeness in the regions covered only by the OGLE-IV fields is
reduced by gaps between CCD detectors of the OGLE-IV mosaic camera. We
estimate that about 7\% of RR~Lyr stars fell into these gaps. On the
other hand, some RR~Lyr stars were detected twice, in the overlapping
regions between adjacent OGLE-IV fields. We used such objects to
estimate the completeness in the area covered by the CCD detectors
(outside the gaps). We checked how many variables from our list could be
potentially detected in the neighboring fields and we compared this with
the number of independently identified objects. Assuming that the
minimum number of data points in the light curves must be larger than
100, in total 460 RR~Lyr stars from our catalog were recorded in the
OGLE-IV databases twice, so we had an opportunity to identify 920
counterparts. We independently detected 864 of them, which gives the
general completeness equal to 94\%.

We expect that the completeness will be larger for RRab stars, due to
their characteristic, asymmetric light curves. Indeed, using the same
method we estimate that our collection of RRab stars is complete to
97\%, while for RRc stars it is only 84\%. Many RRc stars, especially
fainter ones, were not recognized as pulsators due to their nearly
sinusoidal light curves. We also estimated how the completeness depends
on the brightness of stars and the number of epochs. For RR~Lyr stars
fainter then $I=19$~mag the completeness drops down to 82\%, while for
variables brighter than $I=17$~mag it is above 95\%. When the number of
data points in the light curves is below 100, but larger than 50, the
completeness decreases to 77\%.

\Section{Discussion}

\Subsection{RR~Lyr Stars in Globular Clusters}

More than 300 RR~Lyr stars in our collection are plausible members of
globular clusters. Cluster variables (the historical name of RR~Lyr
stars) are not only used as distance indicators, but they also play an
important role in unraveling the early history of globular clusters. The
mere presence of RR~Lyr stars in the cluster suggests that its age is at
least 10~Gyr. Furthermore, the distribution of pulsation periods and the
morphology of the light curves of RR~Lyr stars can be used to estimate
the metallicity of the cluster without spectroscopic measurements.

\MakeTableee{
l@{\hspace{8pt}} c@{\hspace{6pt}} c@{\hspace{6pt}} c@{\hspace{8pt}}
c@{\hspace{6pt}} c@{\hspace{6pt}}}{12.5cm}{Globular clusters containing RR~Lyr star candidates}{\hline \noalign{\vskip3pt}
\multicolumn{1}{c}{Cluster name} & RA & Dec & Cluster & $N_{\rm RR}$ & $N_{\rm fieldRR}$ \\
  & (J2000) & (J2000) & radius [\arcm] & &
 (estimated) \\
\noalign{\vskip3pt}
\hline
\noalign{\vskip3pt}
 NGC~6304 & 17\uph14\upm32\ups & $-29\arcd27\arcm44\arcs$ & 4.0 &  5 &  2.0 \\
 NGC~6355 & 17\uph23\upm59\ups & $-26\arcd21\arcm13\arcs$ & 2.1 &  5 &  0.3 \\
 Terzan~1 & 17\uph35\upm48\ups & $-30\arcd28\arcm11\arcs$ & 1.2 &  9 &  2.0 \\
 NGC~6401 & 17\uph38\upm37\ups & $-23\arcd54\arcm32\arcs$ & 2.4 & 30 &  1.8 \\
 NGC~6441 & 17\uph50\upm13\ups & $-37\arcd03\arcm04\arcs$ & 4.8 & 44 &  2.8 \\
 NGC~6453 & 17\uph50\upm52\ups & $-34\arcd35\arcm55\arcs$ & 3.8 & 10 &  2.5 \\
 Djorg~2  & 18\uph01\upm49\ups & $-27\arcd49\arcm33\arcs$ & 5.0 & 17 & 11.0 \\
 Terzan~10& 18\uph02\upm57\ups & $-26\arcd04\arcm00\arcs$ & 0.8 &  3 &  1.0 \\
 NGC~6522 & 18\uph03\upm34\ups & $-30\arcd02\arcm02\arcs$ & 4.7 & 17 &  6.5 \\
 NGC~6540 & 18\uph06\upm09\ups & $-27\arcd45\arcm55\arcs$ & 0.8 &  3 &  0.5 \\
 NGC~6544 & 18\uph07\upm21\ups & $-24\arcd59\arcm51\arcs$ & 4.6 &  6 &  2.2 \\
 NGC~6558 & 18\uph10\upm18\ups & $-31\arcd45\arcm49\arcs$ & 2.1 &  7 &  0.3 \\
 NGC~6569 & 18\uph13\upm39\ups & $-31\arcd49\arcm35\arcs$ & 3.2 & 21 &  0.8 \\
 M22 (NGC~6656) & 18\uph36\upm24\ups & $-23\arcd54\arcm12\arcs$ & 16.0 & 26 &  4.5 \\
 M54 (NGC~6715) & 18\uph55\upm03\ups & $-30\arcd28\arcm42\arcs$ & 6.0 & 128 &  6.5 \\
\noalign{\vskip3pt}
\hline}

In S11 we listed seven globular clusters that likely host from 3 to 43
RR~Lyr stars. In the present work we increase this list to 15 globular
clusters (Table~2). In column~5 of Table~2 we provide the number of RR~Lyr
stars lying within one cluster angular radius from the cluster
center\footnote{The list of the Milky Way globular clusters is
available at the web page\\ {\it http://messier.obspm.fr/xtra/supp/mw\_gc.html}}.
In the last column we also provide the estimated number of field RR~Lyr stars
that fall by chance inside the area outlined by the cluster radii (we
counted RR~Lyr stars lying in the rings from 1.5 to 2.5 radii from the
cluster centers and rescaled the number of detected stars to the area
occupied by the clusters). It should be, however, stressed that our RR
Lyr stars are only candidates for cluster members, selected based on
pure statistical arguments. Spectroscopic and/or astrometric follow-up
observations of these candidates should provide the ultimate
confirmation of their cluster membership. In the FTP site we provide
the file {\sf gc.dat} with the list of RR~Lyr stars located within the
radii of the globular clusters.

As can be seen in Table~2, globular clusters Terzan~10 and NGC~6540 host
only three RR~Lyr star candidates each and it is possible that all of
them are field variables. In the highly-extincted globular cluster
Terzan~1 we found nine RR~Lyr star candidates of which seven objects may
belong to the cluster. To our knowledge, these are the first pulsating
stars found along the cluster's line-of-sight. Most of these variables have
{\it V}-band apparent magnitudes shifted below our detection limit due
to high interstellar extinction.

The largest sample of RR~Lyr stars (128) was found in the globular
cluster M54 (NGC~6715), that resides at the center of the Sgr
dSph. Most of these variables were already discovered by Rosino and
Nobili (1959), Layden and Sarajedini (2000) and Sollima \etal (2010),
however, we also discovered over a dozen new RR~Lyr stars in M54,
including the first known RRd star in this cluster:
OGLE-BLG-RRLYR-37594. Another newly detected RR~Lyr star located close
to the center of M54 -- OGLE-BLG-RRLYR-37582 -- exhibits two
periodicities: 0.508058~d and 0.623940~d. After separation of these two
modes, it turned out that both periods are associated with the
fundamental-mode pulsations. Thus, we have discovered two unresolved
RRab star lying along the same line-of-sight.

\begin{figure}[t]
\includegraphics[width=12.7cm]{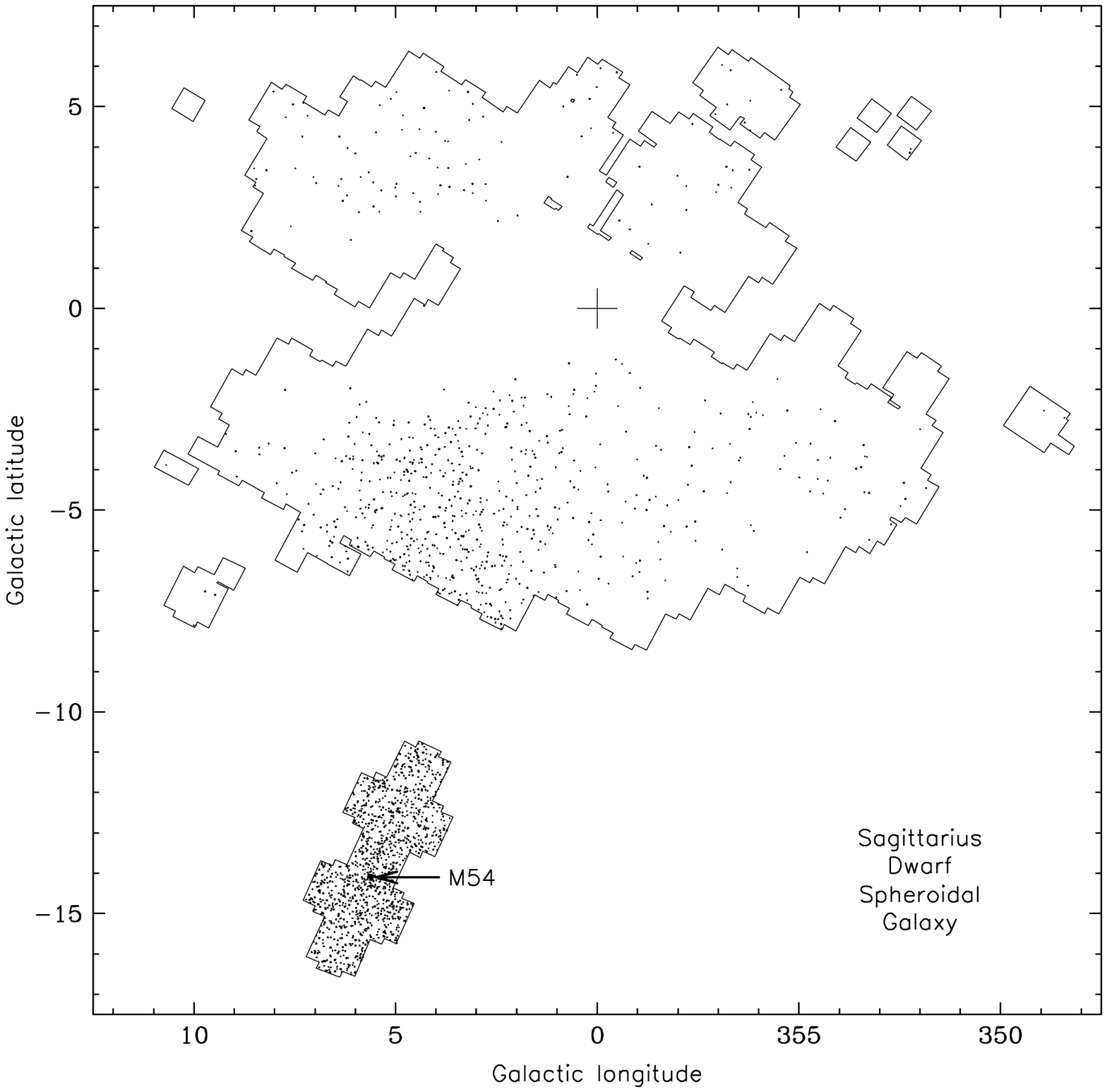}
\FigCap{Spatial distribution of RR~Lyr stars in the Sagittarius Dwarf
Spheroidal Galaxy. The arrow indicates the position of the globular cluster
M54 at the center of the galaxy.}
\end{figure}

\Subsection{RR~Lyr Stars in the Sagittarius Dwarf Spheroidal Galaxy}
More than 2000 RR~Lyr stars from our sample are located far behind the
Galactic bulge. Most of them belong to the Sgr dSph -- a satellite of
the Milky Way that is currently being disrupted by the tidal forces of our
Galaxy. Actually, seven OGLE-IV fields -- BLG705 to BLG711 (the fields
located at the Galactic latitudes $b<-10^{\circ}$) -- were selected
primarily for observations of the central regions of this dwarf galaxy.
Indeed, over 80\% of RR~Lyr stars detected in these fields (about 1300
out of 1600 objects) are members of the Sgr dSph, including 128 cluster
variables in M54. 

Obviously, RR~Lyr stars behind the bulge are affected by different
interstellar extinction, depending on their position in the sky. We
separated RR~Lyr stars behind the bulge using the same arbitrary
criterion as in S11: $I>1.2(V-I)+16.2$~mag. The spatial distribution of
RR~Lyr stars in the Sgr dSph is shown in Fig.~3. A pronounced gradient
in the number of stars is clearly visible.

\begin{figure}[t]
\includegraphics[width=12.7cm]{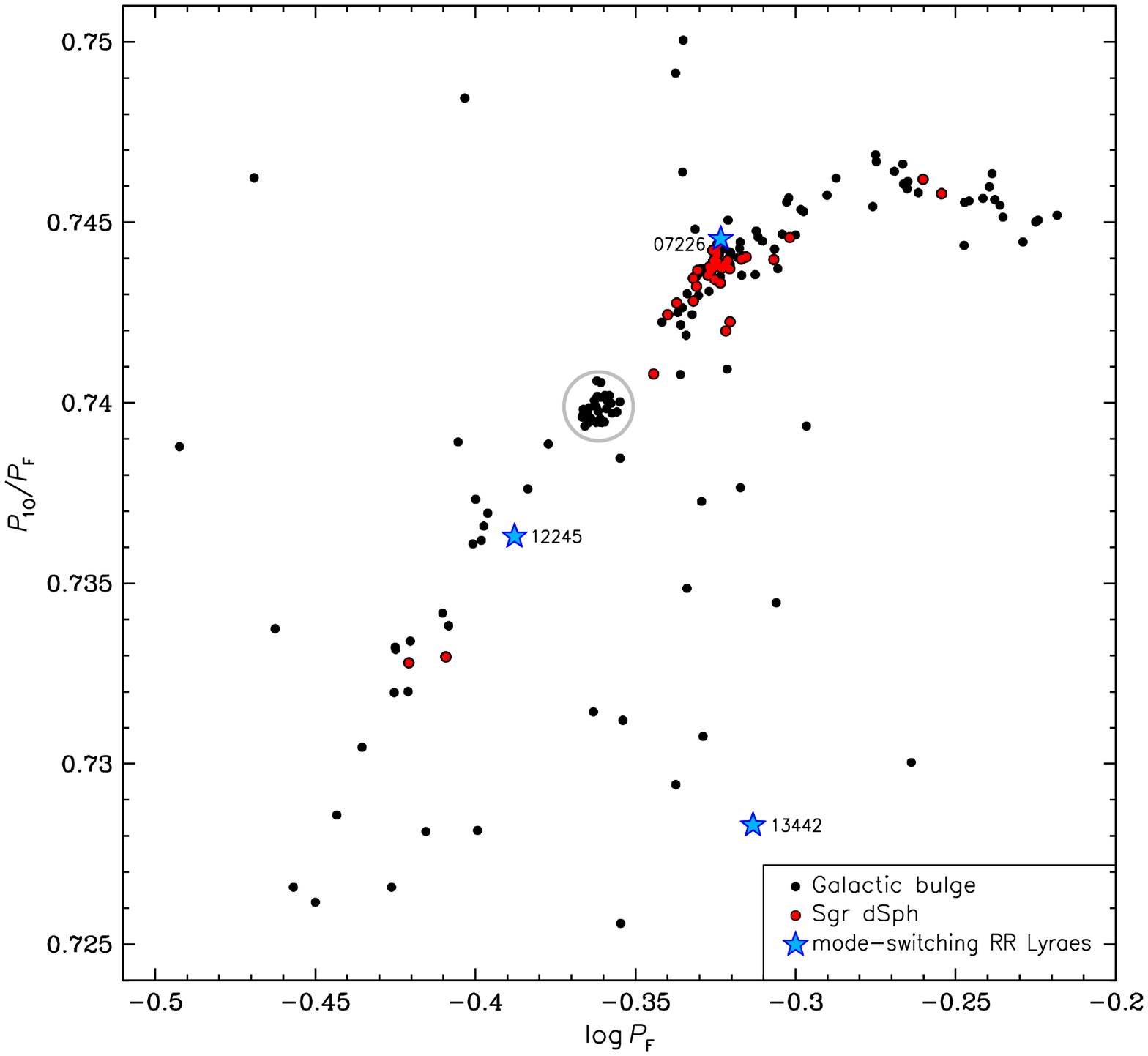}
\FigCap{Petersen diagram for RRd stars in our collection. Black points
indicate objects from the Galactic bulge, red points represent variables
from the Sgr dSph, blue stars show positions of three RR~Lyr stars that
switched pulsation modes. Large gray circle marks the position of a group of
28 RRd stars with $\log{P_\mathrm{F}}\sim-0.36$ and $P_{\mathrm{1O}}/P_{\mathrm{F}}\sim0.740$.}
\end{figure}

\Subsection{Double-mode RR~Lyr Stars}

Double-mode RR~Lyr variables (RRd stars) are very rare objects in the
Galactic bulge, but they exhibit unique features, different than RRd stars
known in any other stellar environments. The number of RRd variables in our
collection almost doubled, compared to our previous study (S11). The
Petersen diagram for this sample is shown in Fig.~4. The extended sample
confirms that some RRd stars in the bulge have exceptionally small
$P_{\mathrm{1O}}/P_{\mathrm{F}}$ period ratios, down to 0.726, which can be
explained by high metallicity values in these objects (S11).

Our sample of RR~Lyr stars also contains a number of double-periodic
variables with the period ratios lower than the above limit. At least
seven of these objects may be RRd stars with the period ratios, between
0.69 and 0.72. The secondary periods of double-periodic RR~Lyr stars are
given in the Remarks of the catalog.

\begin{figure}[t]
\includegraphics[width=12.7cm]{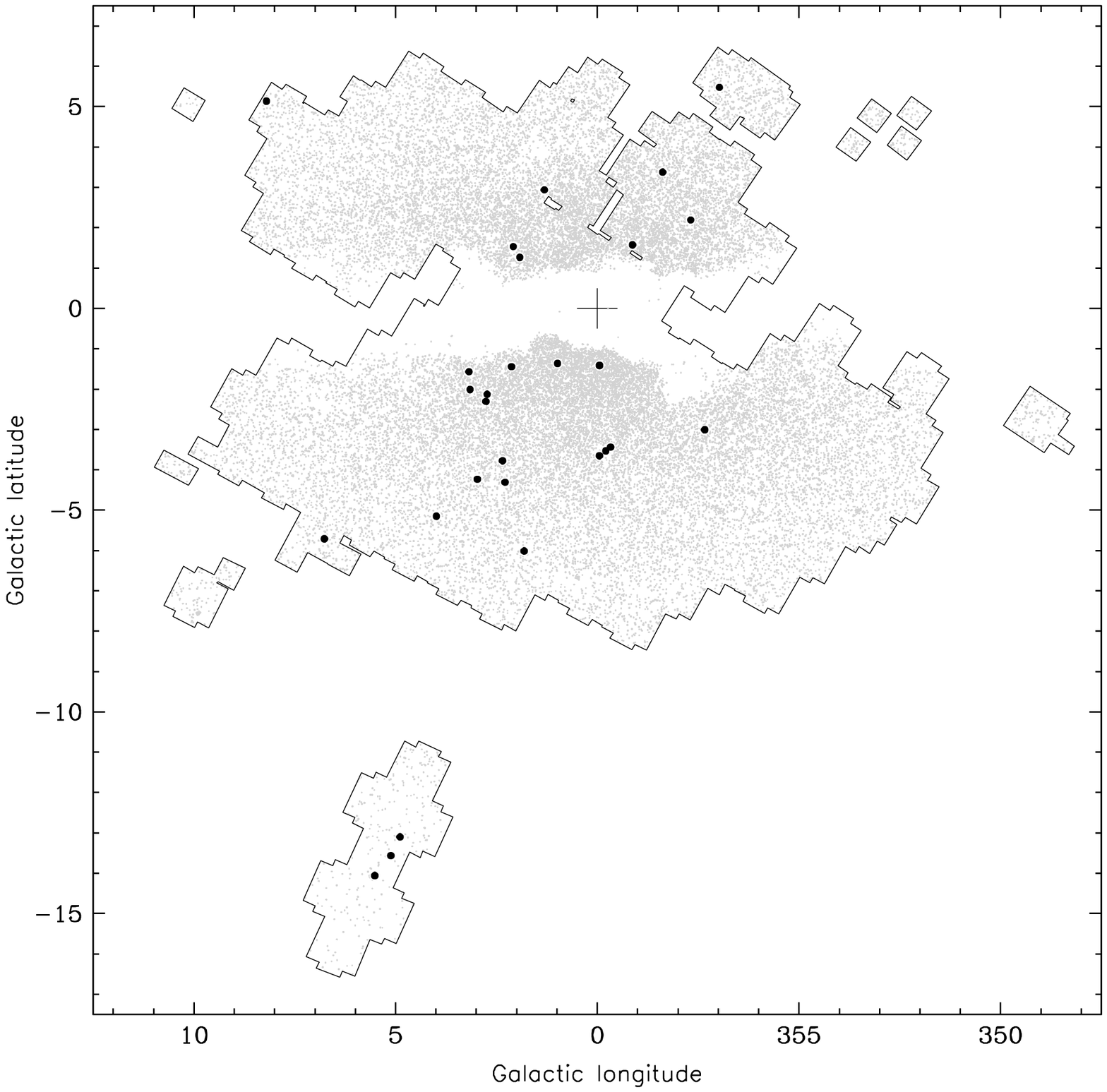}
\FigCap{Spatial distribution of RRd stars encircled in Fig.~4 (black
points). Gray points show the positions of the remaining RR~Lyr stars from
the Galactic bulge (the Sgr dSph variables have been removed).}
\end{figure}

In S11 we noticed a group of 16 RRd stars with a very narrow range of
periods ($-0.366<\log{P_\mathrm{F}}<-0.354$) and period ratios
($0.739<P_{\mathrm{1O}}/P_\mathrm{F}< 0.741$). After adding
OGLE-IV-discovered objects this group increased to 28 stars, which
constitutes over 20\% of all RRd stars detected in the Galactic bulge.
In Fig.~4 these stars are marked with a gray circle. Fig.~5 presents the
position of these double-mode pulsators in the sky. It seems that most
of these objects are associated with a stellar stream that nearly vertically
crosses the bulge. We performed a two dimensional Kolmogorov-Smirnov
test (Fasano and Franceschini 1987) in the $(l,b)$ plane. There is only
a 1.1\% chance that our group of 28 RRd stars and all other bulge RR~Lyr stars
in our collection are drawn from the same general population. Thus, we
suspect that this well-defined group of RRd stars is a relic of a
stellar cluster or a dwarf galaxy disrupted by tidal interactions with
the Milky Way. The stream is similar to the tidal stream of the Sgr
dSph, but it is closer and located roughly at the same distance as the
Galactic center.

\Subsection{RR~Lyrae Stars With Mode Switching}
The long-term extensive OGLE photometry of variable stars offers an
opportunity to study various stellar behaviors, including very rare
phenomena such as mode switching in RR~Lyr stars. Until recently, only one
case of an RR~Lyr star that changed the pulsation mode was known. It was
V79 in the globular cluster M3, which switched from a single-mode RRab
star to a double-mode RRd star in 1992 (Kaluzny \etal 1998) and returned
to a single-mode fundamental-mode pulsation in 2007 (Goranskij 2010).

Soszyñski \etal (2014) presented a second example of a mode-switching
RR~Lyr star -- OGLE-BLG-RRLYR-12245 -- that spectacularly changed from a
double- to single-mode pulsation in a one-year time-span. The third object
of that type -- OGLE-LMC-RRLYR-13308 -- was recently reported by Poleski
(2014), who compared EROS-II (Kim \etal 2014) and OGLE photometry of RR~Lyr
stars in the Large Magellanic Cloud (LMC). Drake \etal (2014) announced the
discovery of six mode-changing RR~Lyr stars in the sample observed by the
Catalina survey, however the example of such an object presented in
their Fig.~31 definitely is not an RRc star that switched to an
RRab star. The nature of the behaviors detected by the Catalina survey in
these six RR~Lyr stars remains an open question.

\begin{figure}[p]
\includegraphics[width=12.7cm]{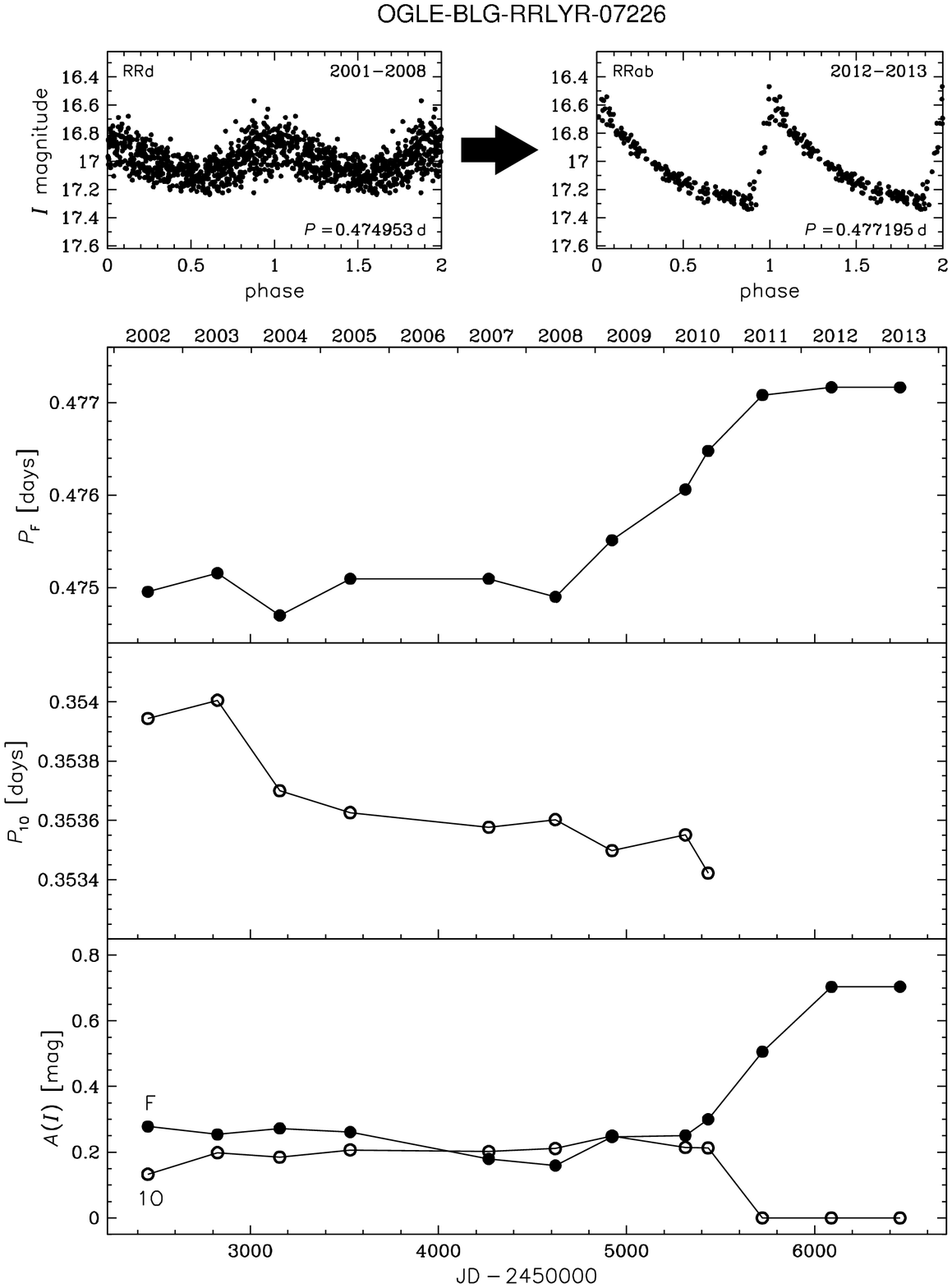}
\vspace*{1pt}
\FigCap{OGLE-BLG-RRLYR-07226 -- an RR~Lyr star that switched from a double-
to single-mode pulsation. {\it Upper panels} show the light curves of
OGLE-BLG-RRLYR-07226 collected in the years 2001--2008 ({\it upper left
panel}) and 2012--1013 ({\it upper right panel}). Both light curves are
folded with the fundamental-mode periods, however note the difference
between both periods. {\it Lower panels} present: changes of the
fundamental-mode period in time, changes of the first-overtone period in
time, changes of the peak-to-peak {\it I}-band amplitudes of both modes.}
\end{figure}

\MakeTableee{
l@{\hspace{8pt}} c@{\hspace{8pt}} c@{\hspace{8pt}} c@{\hspace{8pt}} c@{\hspace{6pt}}}{12.5cm}
{Mode-switching RR~Lyr stars in the Galactic bulge}{\hline \noalign{\vskip3pt}
\multicolumn{1}{c}{ID} & Mode change & $P_F$ (initial) & $P_F$ (final) & $\Delta{P_F}$ \\
  &  & [d] & [d] & [d] \\
\noalign{\vskip3pt}
\hline
\noalign{\vskip3pt}
OGLE-BLG-RRLYR-07226 & \hspace{2.8mm}RRd $\rightarrow$ RRab & 0.474953 & 0.477195 & ~~~0.002242 \\
OGLE-BLG-RRLYR-12245 & \hspace{2.8mm}RRd $\rightarrow$ RRab & 0.409486 & 0.409975 & ~~~0.000489 \\
OGLE-BLG-RRLYR-13442 & RRab $\rightarrow$ RRd & 0.486202 & 0.486074 & $-0.000128$ \\
\noalign{\vskip3pt}
\hline}

We compared the OGLE-III and OGLE-IV light curves of RRd stars in the
bulge and we found additional two RR~Lyr stars that changed their
pulsation modes: OGLE-BLG-RRLYR-07226 that switched from an RRd to RRab
star and OGLE-BLG-RRLYR-13442 that underwent a reverse change. Table~3
summarizes the pulsation characteristics of the three mode-switching
RR~Lyr stars in the bulge. Fig.~6 presents changes of the periods and
amplitudes of OGLE-BLG-RRLYR-07226 during the mode switching.

It seems that all known mode-switching RR~Lyr variables have similar
properties. All these stars switched from a double-mode 1O/F pulsation
(RRd) to a sole fundamental mode (RRab) or {\it vice versa}. In each case
the mode change was a relatively rapid process, lasting about one
year. Every change from a double- to a single-mode pulsation was connected
with the increase of the fundamental-mode period and the reverse change of
modes was associated with the fundamental period decrease.

\Subsection{Unique RR~Lyr Stars}
Our sample of 38\,257 RR~Lyr stars should contain many exceptional
objects. Some of them were presented in the previous Sections. In this
Section we show a few other potentially interesting RR~Lyr stars in our
collection.

OGLE-BLG-RRLYR-14277 underwent an episode of increased luminosity in
2013 due to gravitational microlensing. Its light curve is shown in
Fig.~7. As a microlensing event, OGLE-BLG-RRLYR-14277 was included in
the microlensing alert databases of the large microlensing surveys: MOA
(Bond \etal 2001; as MOA-2013-BLG-300) and OGLE (Udalski 2003; as
OGLE-2013-BLG-0665). Such objects are a powerful tool for constraining
the nature of the gravitational lenses, because the regular variability
of a microlensed star may break the degeneracy between the parallax and
blended light (Wyrzykowski \etal 2006, Assef \etal 2006).

\begin{figure}[htb]
\includegraphics[width=12.7cm]{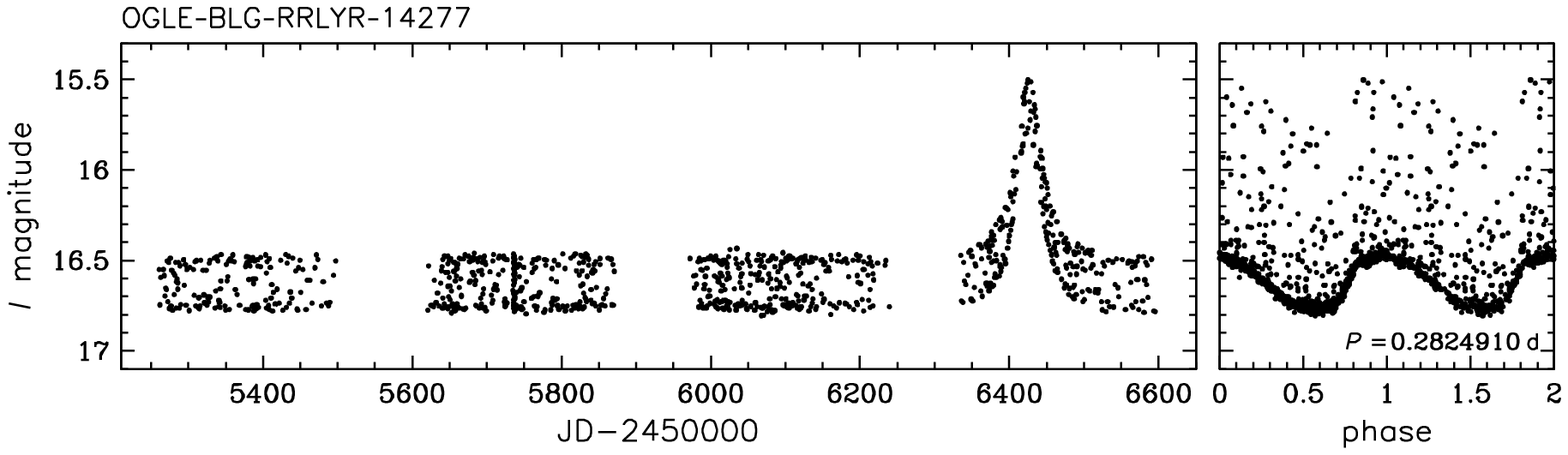}
\FigCap{Light curve of OGLE-BLG-RRLYR-14277 -- an RRc star that experienced
a gravitational microlensing episode. {\it Left panel} presents unfolded
light curve collected in the years 2010--2013. {\it Right panel} show the
same light curve folded with the pulsation period.}
\end{figure}

Many RR~Lyr stars in our sample exhibit large-amplitude Blazhko
modulation. This also applies to several RRd stars. In Fig.~8 we present an
example light curve of such object -- OGLE-BLG-RRLYR-05762. To our
knowledge these are the first known classical RRd stars with Blazhko effect. A
detailed analysis of these stars will be presented in a forthcoming paper
(Smolec \etal 2014, in preparation).

\begin{figure}[htb]
\includegraphics[width=12.7cm]{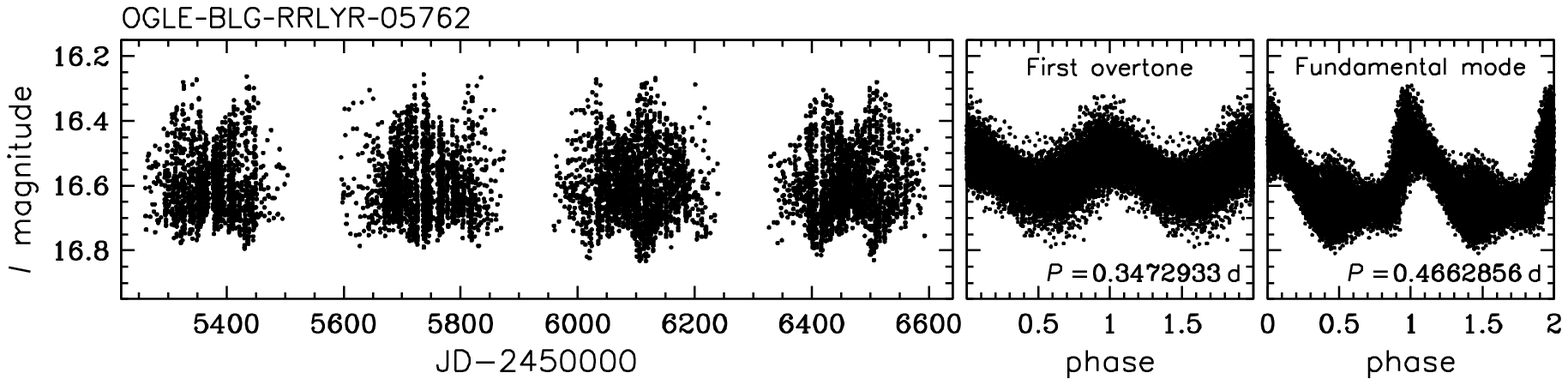}
\FigCap{Light curve of OGLE-BLG-RRLYR-05762 -- an RRd star with the Blazhko
effect. {\it Left panel} presents unfolded light curve collected in the
years 2010--2013. Note the modulations of the pulsation amplitudes. {\it
Right panels} show the pre-whitened light curves of the first-overtone and
fundamental modes.}
\end{figure}

Finally, we draw the attention to other RR~Lyr stars with the Blazhko
effect. The huge number of stars and long-term light curves offer an
opportunity to shed some light on this mysterious phenomenon. As an
example, we present in Fig.~9 an RRab star -- OGLE-BLG-RRLYR-07605. This
is a Blazhko variable which underwent a significant period change in the
years 2007--2011. Since 2011 the light curve of OGLE-BLG-RRLYR-07605 has
been much more stable than in the previous years, and the amplitude of
the Blazhko modulation significantly decreased. This is clearly visible
in the lower panel of Fig.~9, where we present the changes of the
pulsation period together with the changes of the Blazhko amplitudes.
Note the anti-correlation of both quantities. Such a period--Blazhko
amplitude feedback may be an important constraint for the physical
explanation of the Blazhko effect.

\begin{figure}[htb]
\includegraphics[width=12.7cm]{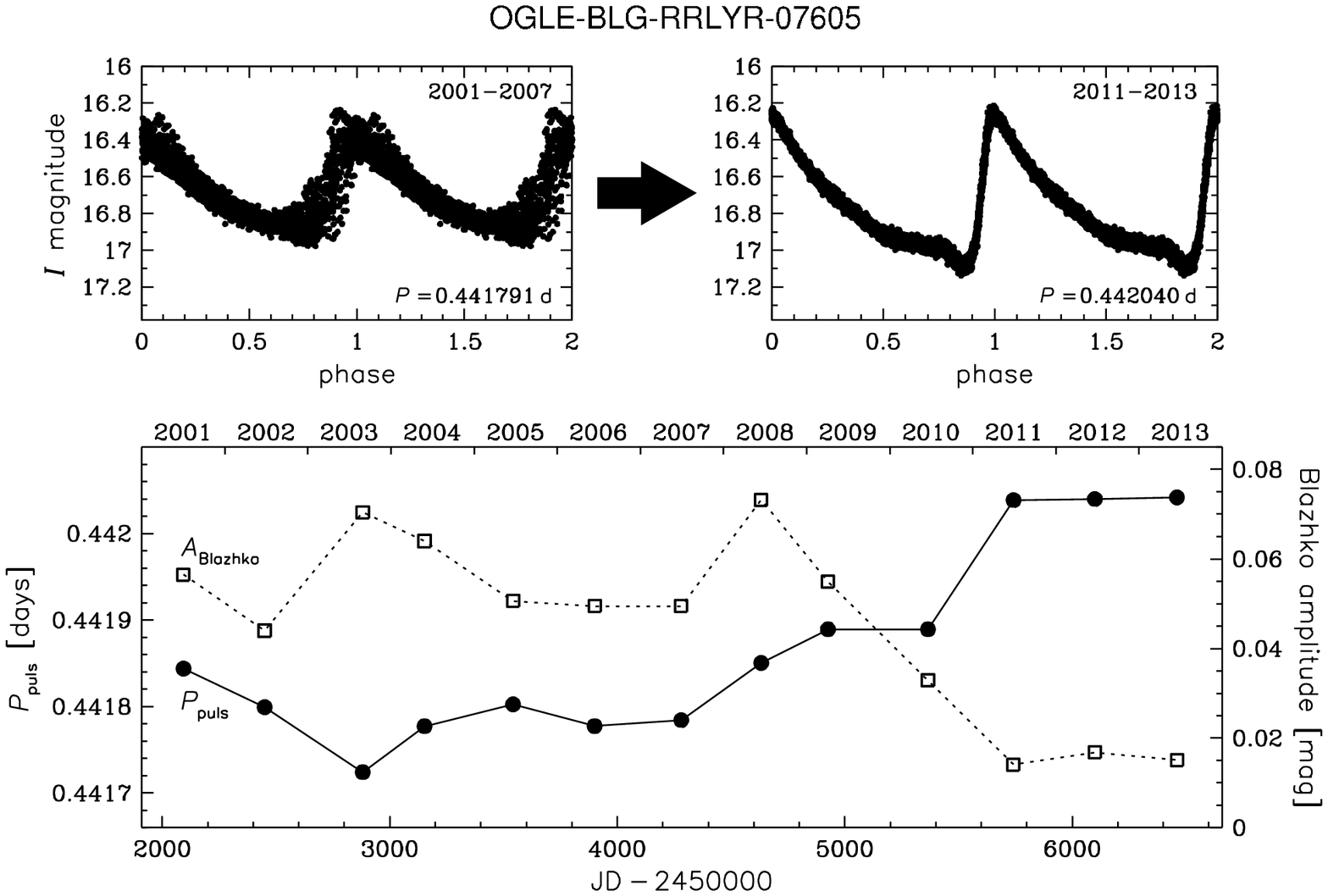}
\FigCap{OGLE-BLG-RRLYR-07605 -- a Blazhko period-changing RRab star. {\it
Upper panels} present light curves of OGLE-BLG-RRLYR-07605 collected in the
years 2001--2007 ({\it upper left panel}) and 2011--2013 ({\it upper right
panel}). {\it Lower panel} shows changes of the pulsation period in time
(filled circles and solid line) and changes of the Blazhko amplitudes
(empty squares and dotted line).}
\end{figure}

\Section{Conclusions}

In this paper, we have presented the most complete picture of the central parts
of the Galactic bulge probed {\it via} RR~Lyr variable stars -- the
OGLE Collection of RR~Lyr variables in the Galactic center. Compared
to our previous study (S11), the OGLE sample of RR~Lyr stars has grown
by a factor of 2.3 in the area of the sky larger by a factor of 2.6.
With 21\,453 RR~Lyr stars identified from the OGLE-IV data, only 1484
objects (7\%) have been cataloged so far in the International Variable Star
Index. In total, the OGLE Collection of the Galactic bulge RR~Lyr
stars contains now 38\,257 objects.

The list of possible astrophysical applications of the OGLE RR~Lyr
collection is very long. For example, it can be used for distance
determinations and mapping the 3-D structures of the bulge and Sgr dSph,
examining the early history of the star formation in the Galaxy,
exploring properties of globular clusters, tracing metallicity
gradients or mapping the interstellar extinction toward the Milky Way
center -- to mention just the basic ones. A huge number of RR~Lyr stars
with long-term light curves is also an ideal laboratory to study various
aspects of stellar pulsations: multi-mode radial and non-radial
oscillations, mode switching, period changes, and the Blazhko effect.

The OGLE picture of the Galactic bulge probed with RR~Lyr stars will
be significantly extended in the coming years. The OGLE-IV project has
recently started another large scale program -- the OGLE Galaxy Variability
Survey which will photometrically cover practically entire Galactic
bulge -- additional 550 square degrees in the sky -- and a fair fraction of
the Galactic disk.

\Acknow{We are grateful to Z.~Ko³aczkowski and A.~Schwarzenberg-Czerny for
providing software which enabled us to prepare this study.

The OGLE project has received funding from the European Research Council
under the European Community's Seventh Framework Programme
(FP7/2007-2013)/ERC grant agreement no. 246678 to AU. This work has been
supported by the Polish Ministry of Science and Higher Education through
the program ``Ideas Plus'' award No. IdP2012 000162.}

\end{document}